\documentclass[letterpaper,11pt]{article} 
\date{Draft. Please do not circulate or cite.}
\usepackage[margin=1in]{geometry}
\usepackage[numbers]{natbib}
\newcommand{\shortcitep}[1]{[\citeyear{#1}]}
\date{Draft. Do not circulate or cite}

\usepackage{url}  
\usepackage{graphicx}  
\usepackage{float}

\usepackage{tikz}
\usetikzlibrary{positioning}
\usetikzlibrary{calc}
\usepackage{pgfplots, pgfplotstable}

\usepackage{amsmath}
\usepackage{booktabs}
\usepackage{multirow}
\usepackage{float}
\usepackage{color}
\usepackage{array}
\usepackage{url}
\usepackage{verbatim}

\usepackage{xspace}
\newcommand{\etal}{{et al.\@\xspace}}

\newcommand{\fsl}{\textsl}

\newcommand{\mps}[1]{}
\newcommand{\ks}[1]{}
\newcommand{\cwh}[1]{}

\pgfplotsset{width=\linewidth,compat=1.8}

\begin{document}

\title{Framing Matters:\\ Predicting Framing Changes and Legislation from Topic News Cycles}

\author{Karthik Sheshadri, Chung-Wei Hang, Munindar P.~Singh}

\maketitle
\thispagestyle{plain}
\pagestyle{plain}

\begin{abstract}
News has traditionally been well researched, with studies ranging from sentiment analysis to event detection and topic tracking. We extend the focus to two surprisingly under-researched aspects of news: \emph{framing} and \emph{predictive utility}. We demonstrate that framing influences public opinion and behavior, and present a simple entropic algorithm to characterize and detect framing changes. We introduce a dataset of news topics with framing changes, harvested from manual surveys in previous research. Our approach achieves an F-measure of $F_1=0.96$ on our data, whereas dynamic topic modeling returns $F_1=0.1$. We also establish that news has \emph{predictive utility}, by showing that legislation in topics of current interest can be foreshadowed and predicted from news patterns.
\end{abstract}

\section{Introduction}

News is known to be a significant driver of public perception \citep{jw10,gunther1998persuasive,mutz1997reading}. In particular, news impacts product development and user concern. Examples include the reaction to negative press about the Facebook News Feed launch \citep{hoadley2010privacy}, and the end of the Quora ``Views'' feature \citep{quora}. News also appears to influence legislation \citep[note 52]{art29drones,evil}.

Not surprisingly, news has garnered much research interest in computer science, with studies addressing news sentiment \citep{1,ns1}, event detection \cite{twitter,ed1}, and topic detection and tracking (TDT) \cite{tdt2,tdt1}.

However, existing literature surprisingly ignores an important aspect of news analysis, namely, \emph{framing}. Framing theory \citep{framing,ft} suggests that how a topic is presented to the audience (called ``the frame'') influences the choices people make about how to process that information. The central premise of framing theory \citep{framing} is that since an issue can be viewed from a variety of perspectives and be construed as having varying implications, the manner in which it is presented influences public reaction.

In general, understanding news framing can be a crucial component of decision-support in a corporate and regulatory setting. To illustrate this, we present a real-life example of the influence of framing on public perception and legislation. In 2011, security vulnerabilities in Facebook's use of HTML5 allowed third-party applications to steal personal data from approximately 59 million users \citep{zynga}. The framing of news on the topic ``Markup Languages'' changed from a neutral narrative to one focusing on personal privacy. Revenues of Merix Games, Wimi5, and other HTML companies declined to the tune of four million dollars over the course of 2012 \citep{sue}. We posit that the decline was caused by negative coverage of news about HTML5 following the data leak (see Figure~\ref{fig-html5}). In 2013, the Personal Data Protection and Breach Accountability Act was promulgated in Congress \citep{pdc}, under which Facebook was sued \citep{sue}.

Accordingly, we address three problems pertaining to framing:

\subsubsection{Detecting Framing Changes}
Since news is an important channel for public communication, the framing of news influences public perception. In particular, \emph{framing changes} have been shown to cause changes in public opinion \citep{smoking,jw10,obesity,1}.

Efforts to estimate changes in framing of a given news topic have hitherto been restricted to post facto studies \citep{smoking,lgbt,jw10,obesity,pew} that require considerable human effort to survey and analyze data. Given the fast pace of cause-and-effect news cycles (see Section on \emph{Discriminative Framing and News Cycles}), it is useful to be able to automatically detect framing changes in an ongoing news cycle. 

\subsubsection{Understanding Public Perception}
Our work builds on recent efforts to understand the drivers of public perception. Whereas companies, regulators, and Government organizations invest considerable manual time and effort into surveying and measuring public perception \citep{public1,public}, self-reported data has proven to be unreliable as a measure \citep{per,sr,pp}. Recent research in sociology has attempted to leverage news as an alternative data source \citep{newsperception,roberts1990news,1}.

Our work confirms that news is a powerful influence and indicator of public opinion, builds on existing work by automating survey-based approaches, presents a data-driven model for news cycle prediction, and quantitatively demonstrates its practical applicability with a case study of legislation, by developing a predictive relationship between news framing and legislative activity.

\subsubsection{Predicting Legislative Changes}

Existing literature in Topic Detection and Tracking \citep{tdt2,tdt1} and event detection only addresses \emph{post facto} inference from news, but stops short of making any predictions about public or legislative reaction. However, ensuring compliance with latest legislation is time and effort intensive for most organizations, and suboptimal compliance monitoring by Internet companies \citep{mm,sc} has resulted in multi-million dollar losses on multiple previous occasions. Since there could be complex exogenous factors that drive legislative interest in a topic, it is hard to anticipate legislation and adapt policy. 

We attempt to address this shortcoming by building a predictive model of legislative activity from news patterns. We present surprising results that demonstrate powerful predictive relationships between topic news patterns and legislation. 

The main contributions of this paper are the following: 

\begin{description}
\item[Detecting Framing Changes] We present what is to our knowledge the first algorithm (see Section on \emph{Changes in Framing}) to quantify changes in framing between two temporally separated news corpora, and threshold changes as significant or insignificant. Our algorithm outperforms Dynamic Latent Dirichlet Allocation (DLDA) \citep{dlda} by a factor of $80\%$ (in terms of $F_1$ measure) on our data.

\item[Data-Driven News Cycle Prediction] We present a data-driven approach (Section on \emph{Discriminative Framing}) that models the relationship between features of a news cycle and likely public reaction. Our model enables us to identify the current ``state'' of a news cycle, its likely duration, and probable public perception change.

\item[Predicting Legislative Activity] We present a system for predicting the likelihood of legislation being enacted in a given topic, based on the corpus of news publishing within it. Our system achieves an overall $F_1$ measure of 0.96 our data set.

\end{description}

We restrict our analysis to \emph{online text-based} news, and focus on the analysis of news patterns in the United States and the United Kingdom.

\section{Related Work}

Entman \shortcitep{entman} defines framing as the ``selection [of] some aspects of a perceived reality to make them more salient in a communicating text, in such a way as to promote a particular problem definition, causal interpretation, moral evaluation, and/or treatment recommendation ''. In Section~\emph{Changes in Framing}, we present an n-gram based approach to the estimation of framing, and an entropy-based approach to framing changes that draws motivation from \citep{entman}.

It is worthwhile to distinguish our study of framing from the problem of event detection, which refers to detecting localized events in a streaming corpus of news. Event detection often relies on bursts in specific n-gram frequency in a \emph{general} news corpus. Topic modeling approaches such as LDA \citep{lda} and Dynamic LDA (DLDA) \citep{dlda} are often used for this purpose. However, within a given news topic, many defining n-grams such as \fsl{smoking} in a smoking corpus remain consistent across events. For example, almost 50\% of DLDA keywords in the example from \emph{Science} \citep{dlda} are identical. Consequently, such keywords and n-grams do not reflect \emph{changes} in language brought about by the associated events. Further, not all events automatically result in a framing change; for instance, many demonstrations in favor of lesbian, gay, bisexual, and transgender (LGBT) rights between 1990 and 2000 (see Section on \emph{Changes in Framing}) did not succeed in altering the overall news framing until the early 2000s.


A related research area is of Topic Detection and Tracking (TDT) \citep{tdt1,tdt2,BHS,BED}. TDT addresses the problem of discovering unsupervised topics from a stream of news stories. It works by clustering similar documents together to form topics. Whereas TDT
could form a baseline approach to identify news documents relevant to a specific topic, it cannot be used to detect framing changes \emph{within} a given topic. 

\section{Datasets}
\subsection{Data Sources}
We examine news articles gathered from The New York Times (NYT) \shortcitep{nytapi} and the Guardian \shortcitep{guardianapi}.
In addition to the large volume of relevant news made available by these two publications (see Section on \emph{Data Sources}), our choice is motivated by their well-documented influence on public attitudes and perception \citep{nyt1,guardian,nyt3,nyt2,mutz1997reading}. We also note that the NYT has previously been shown to influence legislation \citep{evil}, making it an ideal choice for our study.


\subsection{Topic Datasets}
\label{topic}
We define \emph{topic news} as referring to all news publishing primarily related to a specific topic (see Table~\ref{tab:kappa}, and the section on \emph{Dataset Accuracy}).

Our choice of topics is motivated by ground truth framing changes that were observed and recorded using extensive manual studies from earlier research \citep{smoking,lgbt,jw10,obesity,pew}. 

We follow an iterative data collection procedure to seed and refine our datasets for each topic. We begin with a simple keyword search for the topic of interest, for instance, ``surveillance.''  We then extract a random subset of the articles returned by this search (we used a cardinality of 100 in this paper), and manually code these into relevant and irrelevant articles. We define our period-specific universal set $U_{t_1 , t_2}$ as the set of all articles published by the API during the time period $t_1$ to $t_2$. The period $t_1$--$t_2$ is specifically chosen for each topic given \emph{a priori} knowledge of changes in framing (see Section on \emph{Changes in Framing}). We generate \emph{topic negatives} by mining corpora extracted from the ``quiescent'' periods of topic news cycles as described in the \emph{Discriminative Framing} section. Using the positives and negatives from our manual coding, we train a Random Forest (RF) \citep{rf} classifier, which we use to extract a further $m=min(1000,k)$ positives and hard negatives \citep{hn}, where $k$ is the number of positives (negatives) in our universal set. We use $m$ as our training vector in a new RF to extract all positives in $U_{t_1 , t_2}$ which forms our final topic dataset. We find that this iterative training approach increases dataset precision over a single classifier.

\subsection{Dataset Accuracy}
\label{accurate}
To gain confidence that the articles obtained using the approach described above are relevant to each topic, we manually reviewed samples from each topic dataset. In particular, to estimate the precision we coded random samples of 200 each from the NYT and the Guardian from each topic. Each sample was coded by two people; one person coded both sets. We employ a simple coding scheme whereby an article is said to be relevant to a topic if it could not achieve publication with the topic component removed, for instance, an article is about LGBT rights if the segment of the article \emph{not} concerning LGBT rights could not achieve publication in its own right.

Table~\ref{tab:kappa} shows interrater agreement using Cohen's Kappa \citep{vg05}. 

\begin{table}[!htb]
\centering
\caption{Interrater agreement as Cohen's Kappa.}
\label{tab:kappa}
\begin{tabular}{l r r r }
\toprule
&\multicolumn{2}{c}{Precision}\\
\multicolumn{1}{c}{Topic} &  \multicolumn{1}{c}{Coder~1} & \multicolumn{1}{c}{Coder~2} & \multicolumn{1}{c}{Kappa}\\
\midrule
LGBT rights & 0.98 & 0.98 & 1.00\\

Smoking & 0.96 & 0.98 & 0.85\\

Surveillance & 0.83 & 0.82 & 0.96\\ 

Obesity & 0.91 & 0.88 & 0.82\\

Cyberbullying & 0.86 & 0.93 & 0.87\\

Drones & 0.72 & 0.76 & 0.64\\

HTML5 & 0.98 & 0.98 & 1.00\\
\bottomrule
\end{tabular}
\end{table}

\section{Changes in Framing}
\label{frame}

We detect changes in framing based on computing entropy.

\subsection{Discriminative Framing and Keywords}
\label{key}

We define \emph{discriminative framing} as those aspects of a topic's current framing that distinguish it from the topic's framing at a previous time period. To this end, we adopt the idea of an entropic formulation of discriminative keywords, as proposed by Sheshadri {\etal} \shortcitep{1}.

Below, a corpus $T$ is a set of news articles. 
Specifically, given two disjoint sets of news articles $T_1$ and $T_2$, we identify a set of $k$ $n$-grams that yield the largest Information Gain (IG) in the combined corpus $T=T_1\cup T_2$.
Let $A$ be an article in corpus $T$.  Let $x_i$ represent any of the possible $m$ n-grams in $T$.  Let $S(x_i, T)=\{A\in T|x_i\in{A}\}$ be the set of articles in corpus $T$ in which the n-gram $x_i$ appears. We use a $|T|\times m$ term frequency (TF) matrix representing the corpus to calculate $H$, the information entropy of $T$.

\begin{equation}
IG(T,x_i)=H(T)-\frac{S(x_i, T)}{|T|}H(S(x_i))\\
\end{equation}

Following Entman's formulation, this approach weights n-grams that are specific to a particular corpus more highly than n-grams that are common to both corpora. A quick intuition for the approach is obtained by considering that the unigram ``Snowden" has a high utility in distinguishing Surveillance articles subsequent to 2014 from those prior to 2013, but the unigram ``surveillance'' is common to both corpora and therefore does not.

Since keywords from a particular news corpus distinguish it from others, they may be said to represent the ``state'' or ``concentration'' of news in that corpus.

We represent n-grams in a learned co-occurrence vector space in order to compare similarity. To simplify computation, we conduct Singular Value Decomposition (SVD) on our co-occurrence space and extract the eigenvectors corresponding to the $j=3$  largest eigenvalues. We then compute Word Mover's Distance (WMD) \citep{wmd} between all pairs of keywords in this space, and use their median as a simple numeric threshold for significance. The values of $k=6$ and $j=3$ are experimentally selected to optimize the observed $F_1$ measure. We used unigrams and bigrams, following previous research \citep{1}.

To arrive at a threshold for significance, we used an Expectation Maximization (EM) approach to choose the value (0.15 in this paper) that maximized performance.\cwh{I am not clear about how EM is applied}

Note that aside from experimental optimization of our parameters, the approach we use to threshold framing changes is completely unsupervised and requires no manual labeling of training data (other than the assumption of a high precision dataset, generally obtainable using the procedure described in Section \emph{Topic Datasets}).

\begin{figure*}[!htb]
\centering
\includegraphics[width=\textwidth]{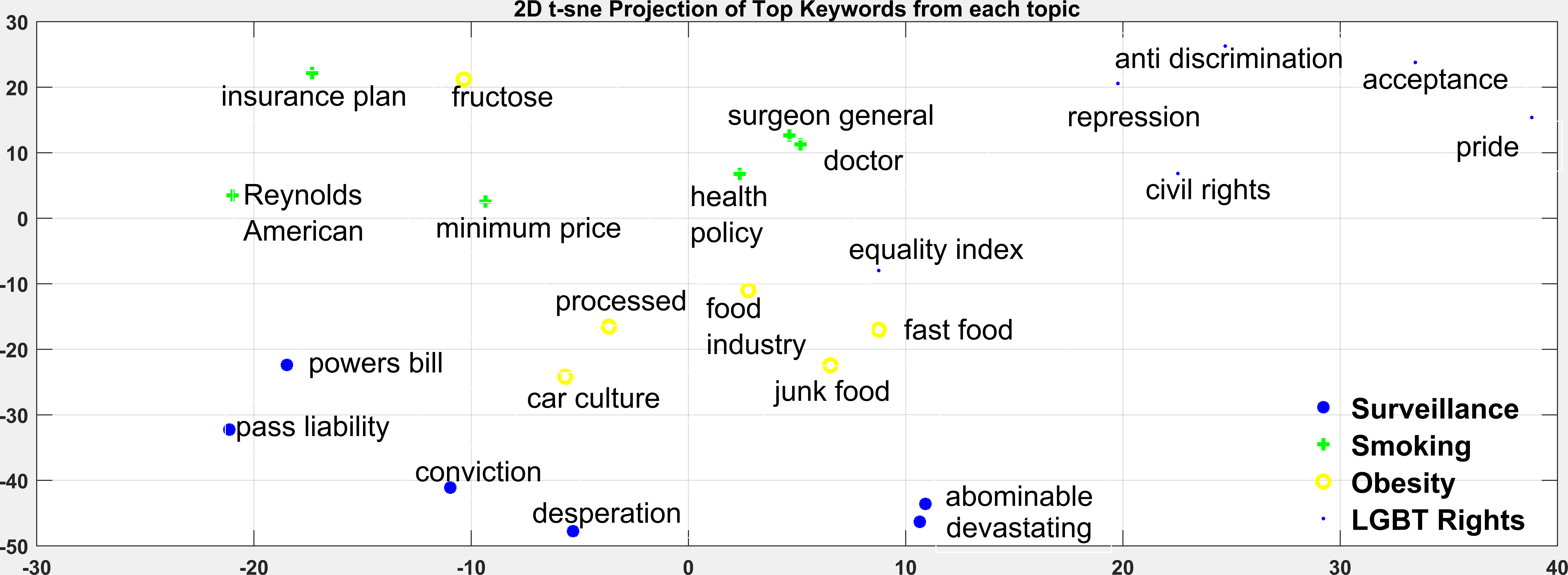}
\caption{A 2D projection of maximum entropy keywords within each of the four framing change positives. Notice that keywords within each topic cluster together, illustrating the framing change.}
\label{fig:tsne}
\end{figure*}

To demonstrate the semantic similarity between top keywords in each topic, we used t-SNE \citep{tsne} to obtain a 2D projection of our learned similarity space. The clusters and separation are depicted in Figure~\ref{fig:tsne}.

\subsection{Smoking}
Cummings \shortcitep{smoking} conducted a study of the framing of smoking related news pre and post 2000, and concluded that framing had changed from portraying smoking as a personal rights issue from 1990--2000 to coverage primarily as a health issue post 2000. To check if our system detects this change, we extracted 2,000 articles from each API, 1,000 each for the two periods 1990--2000 and 2001--2017. 

Table~\ref{smoke} demonstrates our keyword results alongside DLDA results on this dataset. Whereas DLDA results for each period are relatively consistent and do not reflect the ground truth framing change, our approach identifies many keywords that are indicative of the change in framing, for example, the $n$-grams ``surgeon general'', ``doctor'', ``insurance plan'' and so on are reflective of the change. Our mean semantic similarity on this dataset ($0.2$) is thresholded as significant by the EM approach described in the \emph{Discriminative Framing} section. 

\begin{table}[!ht]
\centering
\caption{Comparing our approach with DLDA. The sets of top DLDA keywords for the consecutive periods are almost identical, indicating that framing changes are not identified by DLDA. In contrast, our approach highlights the framing changes by bringing out the keywords that indicate the change of framing across the same two periods.}
\label{smoke}
\resizebox{.9\columnwidth}{!}{%
\begin{tabular}{@{~} l @{~~} p{0.64in} @{~~} p{0.85\columnwidth} @{~}}
\toprule
\multirow{4}{*}{\rotatebox[origin=c]{90}{\parbox[c]{100pt}{\centering Smoking}}}
&DLDA 1990--2000 & cigarette; tobacco; reynolds; advertising sales; philip morris  \\[6pt]

&DLDA 2001--2017 & cigarette; tobacco; reynolds; reynolds american; philip morris; electronic cigarettes\\[6pt]

& DLDA\newline difference & advertising, sales; electronic cigarettes\\[6pt]

&Our\newline difference & doctor; surgeon general; insurance plan; reynolds; health policy; minimum price \\
\midrule
\multirow{4}{*}{\rotatebox[origin=c]{90}{\parbox[c]{110pt}{\centering Surveillance}}}
&DLDA 2003--2013 & surveillance; patient records; cookies; Google; Harry Cayton; communications\\[6pt]

&DLDA 2014--2017 & Freedom Act; phone records; patriot act; Snowden; whistleblower; stingray\\[6pt]

& DLDA\newline difference & surveillance; patient records; cookies; Google; Harry Cayton; Freedom Act; phone records; Patriot Act; Snowden whistleblower; stingray \\[6pt]
 
&Our\newline difference & Snowden; desperation; powers bill; abominable; devastating; pass liability\\
\midrule
\multirow{4}{*}{\rotatebox[origin=c]{90}{\parbox[c]{100pt}{\centering LGBT Rights}}}

&DLDA 1990--2000 & lgbt; gay; gay marriage; conservative; gay rights; bigotry\\[6pt]
&DLDA 2007--2017 & gay; lgbt people; same sex; gay rights; transgender; gay travelers \\[6pt]
&DLDA\newline difference & gay marriage; conservative; bigotry; same sex; transgender; gay travelers\\ 

&Our\newline difference & pride; anti discrimination; acceptance; repression; equality index; civil rights \\
\midrule
\multirow{4}{*}{\rotatebox[origin=c]{90}{\parbox[c]{120pt}{\centering Obesity}}}
&DLDA  & sedentary lifestyle; diet; health warning \\[6pt]

&1990--2000 &  genetic causes; unhealthy; exercise \\[6pt]

&DLDA 2007--2017& obesity; unhealthy diet; genetic; low intensity; fat intake; surgical treatment\\[6pt]

&DLDA\newline difference & sedentary lifestyle; health warning; genetic causes; exercise; unhealthy; obesity; fat intake; surgical treatment\\[6pt]

&Our\newline difference & food industry; car culture; processed; fast food; fructose; junk food \\
\bottomrule
\end{tabular}%
}
\end{table}

\subsection{LGBT Rights}
Reference \citep{jw10} shows that public approval of LGBT rights is at its lowest (within the period surveyed) from 1996--1998 at an average 30\%, and shot up to 46\% in 2007. Further, the paper describes how the framing of LGBT rights has changed from being seen as a morality issue pre 2000, to being seen as an equal rights issue post 2007. We evaluated our approach on the framing of LGBT rights by extracting news from the NYT API during two periods, 1990--2000 (2,332 articles), and 2007--2017 (3,176 articles). We restrict our analysis to NYT, since the study in \citep{jw10} uses a survey of American residents.

We present top keywords in Table~\ref{smoke}. Several of our keywords reflect the framing change, for example, the $n$-grams ``equality index'', ``acceptance'', ``anti discrimination'' and so on are indicative of the change in news framing to an equal rights narrative. 


\subsection{Surveillance}
A Pew Research survey \citep{pew} found that public approval of the NSA surveillance program declined sharply to about 25\% from the earlier $49\%$ in the wake of the Snowden revelations \citep{pew}. Since then, publishing in Surveillance has reflected a trend in public disapproval (manifesting in negative framing) and skepticism of privacy protection \citep{sn}.

We evaluated our approach on this dataset by extracting surveillance articles from both APIs for the period 2003--2017. 

Table~\ref{smoke} depicts the results. Note that whereas DLDA keywords for this dataset do not remain largely unchanged over the two periods, these keywords still do not detect the change in framing to more negative sentiment coverage.

Our keywords exhibit enhanced performance over DLDA on this dataset in two ways. Firstly, mean semantic similarity in our learned co-occurrence space for our keywords (0.16) represents a significant (as determined by manual surveys) framing change according to the EM classifier\cwh{It is not a classifier, is it?}, whereas the corresponding similarity from DLDA (0.09) is thresholded as not significant (contrary to ground truth). Secondly, three out of our top six keywords figure in a standard list of negative sentiment words \citep{breen}, reflecting the more negative coverage of surveillance news subsequent to the Snowden revelations (Figure~\ref{fig-surveillance}) \citep{pew}. 

\subsection{Obesity}
Obesity related news \citep{obesity} framed the issue as primarily one of individual responsibility pre 2001, framing in the last 15 years had in contrast presented the issue as being primarily due to societal and cultural problems. We scraped 2,000 articles from the New York Times (since \citep{obesity} restricts its study to Americans) from 1990--2000 and 2001--2017, respectively. 
Top keywords from our approach alongside DLDA are shown in Table~\ref{smoke}.

\section{Discriminative Framing and News Cycles}
\label{cycle}
In the previous Section, we showed that semantic similarity of discriminative framing serves as a high F-measure classifier for ground truth topic framing changes. In this section, we apply these insights of to develop a theory linking news cycles to public reaction.
News publishing within a specified topic is often driven by significant events \citep{news2,news1}. For example, consider the effect of the Snowden revelations on publishing volume  in Surveillance (Figure~\ref{fig-surveillance}). The number of surveillance news articles increased by nearly 250\% for the year 2014. To confirm that the increase in volume was due to Snowden, a single rater coded all surveillance articles from 2014 into as Snowden or not Snowden. The criterion used was to consider an article Snowden related if it could not achieve publication with the Snowden component removed. We found that 67 of 72 Surveillance articles in 2014 are Snowden related.

We use the fact that event-driven publishing is likely (Figures~\ref{fig-surveillance}, \ref{fig-cyberbullying}, \ref{fig-drones},~\ref{fig-child} and~\ref{fig-html5}) to have high similarity, such as in the Snowden example above. We posit that significant events can thus be said to create high volume, \emph{correlated} publishing within a topic, which is likely to elicit a public response. Following this response, publishing within the topic dies away to a quiescent state in which volume is low and publishing is uncorrelated (Figures~\ref{fig-surveillance}, \ref{fig-cyberbullying}, \ref{fig-drones},~\ref{fig-child} and~\ref{fig-html5}).

Motivated by this framework, we define the following \emph{features} of a news corpus:

\subsubsection{Volume:} The number of articles in a news corpus.

\subsubsection{Mean Sentiment:} The mean polarity \citep{stanford} of the articles in a news corpus.

\mps{I simply don't understand the concepts here. What is correlation? How should we read the figures in this paper? For example, what's going on in Figure 2? Can you step through it in detail? Once you do, we can shorten the description if necessary.}

\subsubsection{Mean Normalized Correlation:} The mean pairwise Pearson correlation \citep{pearson} between each pair of $\binom{n}{2}$ articles (represented by TF vectors) in a corpus, where $n$ is the volume of the corpus.

Whereas earlier work presented evidence that news influences public reaction, it does not address prediction of the nature of that reaction or when it is likely to occur. Our findings suggest that certain features of news, such as volume and correlation (as described above), changes in framing, and sentiment variations (Figure~\ref{fig-surveillance}) present sources for a data driven learning framework which holds the promise of automating predictions of public reaction.  
Further, the aforementioned features also serve as an indicator of \emph{where} in a news cycle a topic currently resides, and how much longer it may endure.

\section{Legislation}

Public reaction to news publishing has been shown to have diverse implications, such as in technology development \citep{quora}, user behavior \citep{frank2012mining}, and regulatory policy \citep{art29drones}.

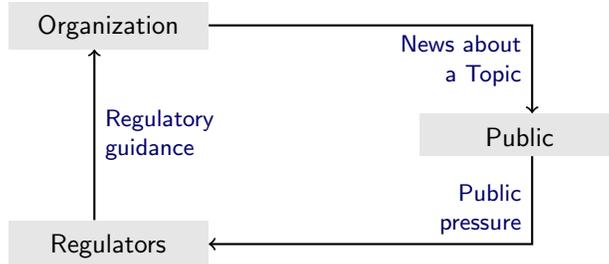
\begin{figure}[!htb]
\centering
\begin{tikzpicture}[%
    obj/.style={fill=gray!20,text height=0.3cm,text centered,text width=2.4cm},
    node distance=1cm and 4cm, auto,font=\small\sffamily]

  \tikzstyle{edge_label}=[blue!40!black,draw=none,font=\footnotesize\sffamily,text width=1in, align=right]
    
    \draw node [obj] (O) {Organization};
    \draw node [below=of O] (E) {}; 
    \draw node [obj, right=of E] (P) {Public};
    \draw node [obj,below=of E] (R) {Regulators};

    \draw [->, thick] (O.east) -| node[edge_label, pos=0.7, left] {News about\\a Topic} (P.60);
    
    \draw [->,thick] (P.300) |- node[edge_label, pos=0.3, left] {Public\\pressure} (R.east);
    
    \draw [->,thick] (R.120)--
        node[edge_label, align=left, right] {Regulatory\\guidance}(R.120|-O.south);
\end{tikzpicture}%
\caption{News as a communication channel for organizational information. Public perceptions about topics in general and organizations in particular are driven by news. Incidents involving organizations are reported by news publications. News reports influence regulators and regulators often encourage organizations to communicate information through the news \mps{don't understand ``encourage organizations to communicate information through the news''---Facebook should get NYT to publish an article about HTML5?}.}
\label{fig:policy}
\end{figure}

We present strong evidence that news-driven public perception changes can influence \emph{legislative activity}. We demonstrate a consistent predictive relationship between the features of news presented in the \emph{Discriminative Framing} Section and the volume of legislative activity (measured by the count of laws debated, enacted, or amended) within a topic.


We predict an annual binary legislative label. For a given news topic in a given year, the label represents a prediction about whether there is likely to be significant legislative activity (based on the historical news pattern in that topic). We use the features mentioned above. Rather than using the raw values of these features, which are unlikely to be predictive in themselves, we use the absolute value of the normalized \emph{annual difference} of each feature, since changes are more likely to determine where in a new cycle a topic currently resides. As an intuitive illustration, consider that a yearly news volume of 80 articles or a mean correlation of 0.2 is not predictive in itself, but an \emph{increased volume} of 80 articles from last year or an \emph{increased correlation} of $20\%$ may be predictive.

Our training data thus consists of temporally separated pairs of values for each feature. For every pair of observations (for every feature), we construct discrete probability distributions and normalize them over each feature, to arrive at a pair of distributions which capture the change pattern exhibited by the feature over legislative years and years with no legislation.

We then use these distributions to construct a joint change distribution of the two observations. This enables us to arrive at the conditional using a simple Bayesian formulation:

Let $(x_1,x_2)$ be a feature vector pair, then:

\begin{equation}
P_f(x_2|x_1)=\frac{P_f(x_1|x_2)}{\sum_{x_2}{P_f(x_1|x_2})}
\end{equation}

We adopt the naive Bayes assumption to arrive at a single estimate from all the features:

\begin{equation}
P(x_2|x_1)=\prod_{f}P_f(x_2|x_1)
\end{equation}

We use a simple binary threshold to evaluate a binary legislative or not legislative label, $P(x_2|x_1)<t$. 

For our experiments, we adopt the Leave One Out (LOO) approach, employing data from all but one class for training and using the remaining class as our testbed.\cwh{We can consider to find a simple baseline to compare against.}

\subsection{Surveillance}
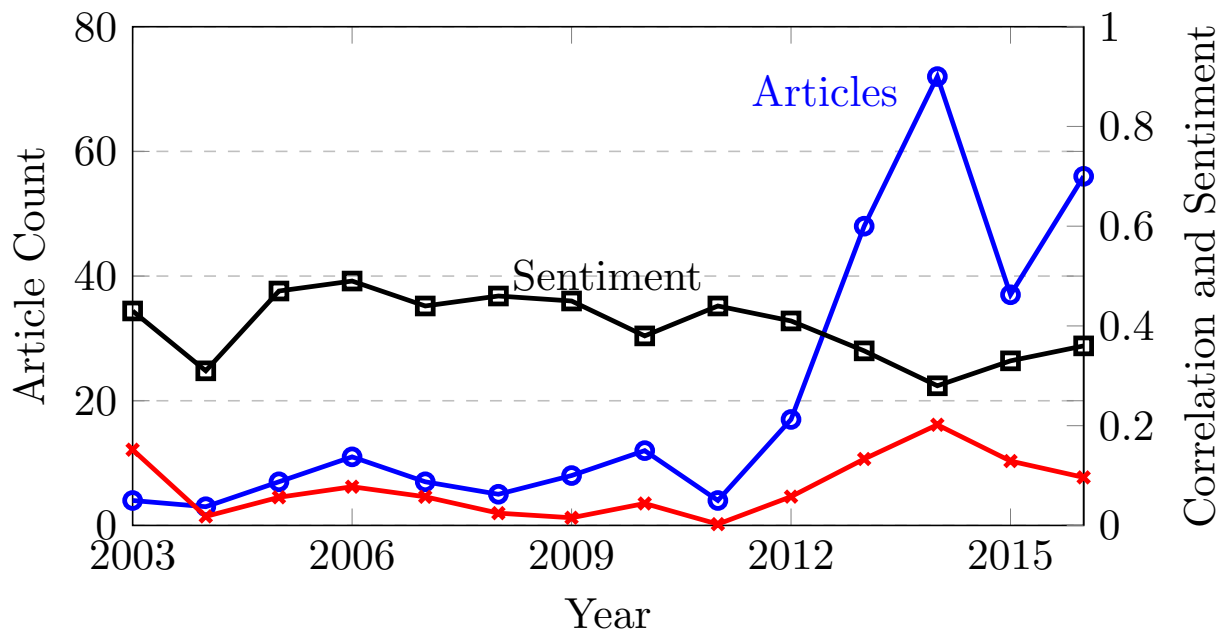
\begin{figure}[!htb]
    \centering
    \resizebox{0.99\columnwidth}{!}{%
\begin{tikzpicture}
        \begin{axis}[
        title={},
        height=6cm,
        width=10cm,
        xlabel={Year},
        ylabel={Article Count},
        xmin=2003, xmax=2016,
        ymin=0, ymax=80,
        xtick={2003,2006,2009,2012,2015},
        ytick={0,20,40,60,80,100},
        legend pos=north west,
        legend style={font=\tiny},
        mark size=2.0pt,
        ymajorgrids=true,
        grid style=dashed,
        xticklabel style={/pgf/number format/.cd, set thousands separator={}},
        ]

        \addplot [very thick,blue,mark=o]
        coordinates {(2003,4) (2004,3) (2005,7) (2006,11) (2007,7) (2008,5) (2009,8) (2010,12) (2011,4) (2012,17) (2013,48) (2014,72) (2015,37) (2016,56)}
        node[above left,pos=0.6] {Articles};

    \end{axis}

   \begin{axis}[
    height=6cm,
        width=10cm,
  axis y line*=right,
  axis x line=none,
xmin=2003, xmax=2016,
  ymin=0, ymax=1,
ytick={0,0.20,0.40,0.60,0.80,1.00},
  ylabel={Correlation and Sentiment}  
]

  \addplot [very thick,red,mark=x]
        coordinates {(2003,0.151819195) (2004,0.01804966519) (2005,0.0564282172339713) (2006,0.0775952248754797) (2007,0.0574131362074208) (2008,0.0246369129971187) (2009,0.0153641043316717) (2010,0.0440065159756945) (2011,0.00257047230208002) (2012,0.0577160220712493) (2013,0.1332121718) (2014,0.2018622475) (2015,0.1290804976) (2016,0.0963360531614937)}
        node[below right,pos=0.65] {Correlation};

\addplot [very thick,black,mark=square]
        coordinates {(2003,0.43) (2004,0.31) (2005,0.47) (2006,0.49) (2007,0.44) (2008,0.46) (2009,0.45) (2010,0.38) (2011,0.44) (2012,0.41) (2013,0.35) (2014,0.28) (2015,0.33) (2016,0.36)}
        node[above=0.1cm,pos=0.5] {Sentiment};

    \end{axis}

\end{tikzpicture}%
}
    \caption{News volume, correlation, and sentiment as predictors of legislation in Surveillance. \mps{The way to draw with multiple Y axes would be to place the addplot commands within a begin-axis to end-axis block. When I moved the sentiment and correlation addplot commands to the second Y axis, it didn't work. (I moved it back.) This is because you seem to have manually altered the correlation values and you have written the sentiment values on a scale other than [0,1]. I suggest you state the correlation and sentiment data as it is (within the [0,1] interval). Then it should print correctly. }}
    \label{fig-surveillance}
\end{figure}

Following the Snowden revelations of June 2013, the USA Freedom Act was introduced in the US Congress in October 2013, and was finally passed into law in 2015. Our Surveillance data captures this period of legislative activity, together with a quiescent period preceding it from 2003 to 2013. Figure~\ref{fig-surveillance} shows the patterns. We obtain an F-measure of 0.93 and an accuracy of 0.93 (13 out of 14) on this dataset.

\subsection{Cyberbullying}

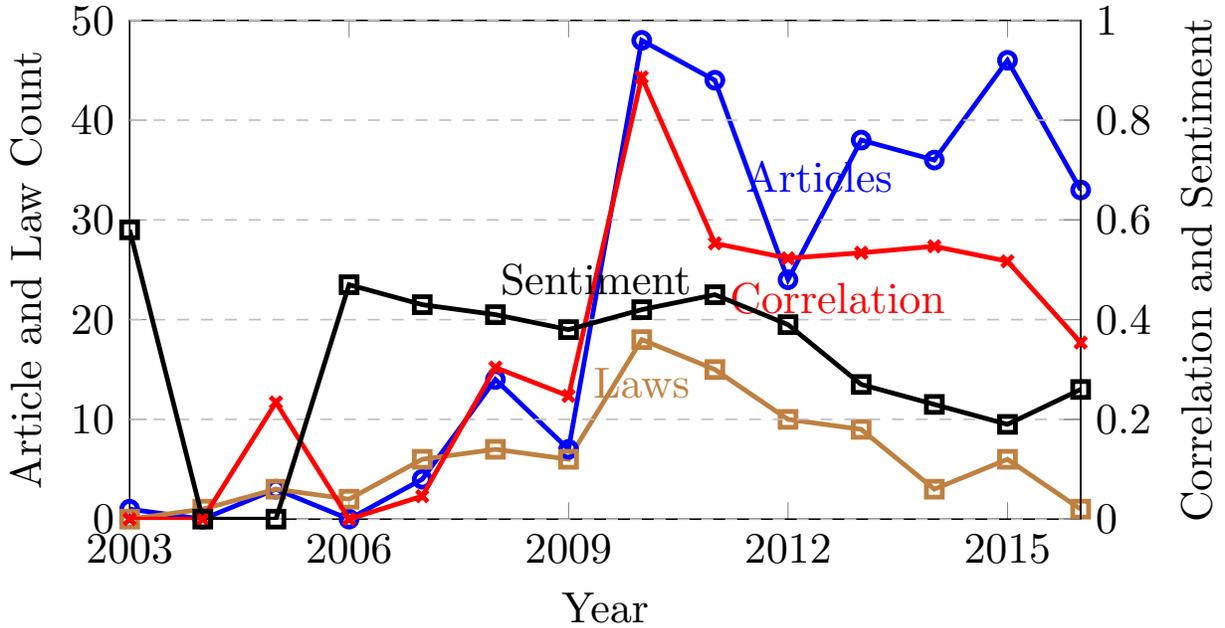
\begin{figure}[!htb]
    \centering
    \resizebox{0.99\columnwidth}{!}{%
\begin{tikzpicture}
        \begin{axis}[
        title={},
        height=6cm,
        width=10cm,
        xlabel={Year},
        ylabel={Article and Law Count},
        xmin=2003, xmax=2016,
        ymin=0, ymax=50,
        xtick={2003,2006,2009,2012,2015},
        ytick={0,10,20,30,40,50},
        legend pos=north west,
        legend style={font=\tiny},
        mark size=2.0pt,
        ymajorgrids=false,
        grid style=dashed,
        xticklabel style={/pgf/number format/.cd, set thousands separator={}},
        ]

        \addplot [very thick,blue,mark=o]
        coordinates {(2003,1) (2004,0) (2005,3) (2006,0) (2007,4) (2008,14) (2009,7) (2010,48) (2011,44) (2012,24) (2013,38) (2014,36) (2015,46) (2016,33)}
        node[above=.1cm,pos=0.75] {Articles};

        \addplot [very thick,brown,mark=square]
        coordinates {(2003,0) (2004,1) (2005,3) (2006,2) (2007,6) (2008,7) (2009,6) (2010,18) (2011,15) (2012,10) (2013,9) (2014,3) (2015,6) (2016,1)}
        node[below=.1cm,pos=0.5] {Laws};

    \end{axis}

    \begin{axis}[
    height=6cm,
        width=10cm,
  axis y line*=right,
  axis x line=none,
 xmin=2003, xmax=2016,
  ymin=0, ymax=1,
ytick={0,0.20,0.40,0.60,0.80,1.00},
 legend pos=north west,
        legend style={font=\tiny},
        mark size=2.0pt,
        ymajorgrids=true,
        grid style=dashed,
  ylabel={Correlation and Sentiment}     
]

      \addplot [very thick,red,mark=x]
        coordinates { (2003,0) (2004,0) (2005,0.234) (2006,0) (2007,0.046) (2008,0.304) (2009,0.247) (2010,0.886) (2011,0.553) (2012,0.523) (2013,0.534) (2014,0.547) (2015,0.517) (2016,0.354)}
        node[below=.1cm,pos=0.75] {Correlation};

         \addplot [very thick,black,mark=square]
        coordinates {(2003,.58) (2004,0) (2005,0) (2006,.47) (2007,.43) (2008,.41) (2009,.38) (2010,.42) (2011,.45) (2012,.39) (2013,.27) (2014,.23) (2015,.19) (2016,.26)}
        node[above=.1cm,pos=0.5] {Sentiment};

    \end{axis}

\end{tikzpicture}%
}
    \caption{News volume and correlation as predictors of legislation in Cyberbullying.}
    \label{fig-cyberbullying}
\end{figure}

Although there are no federal Cyberbullying laws yet, we compiled a comprehensive list of state wise Cyberbullying laws to employ as ground truth. Due to space constraints, we do not enumerate the list here, but provide a few representative entries to illustrate the list. We harvested news articles from 2003 when reports of Cyberbullying began to appear, until 2016 for a total of 375 articles. Figure~\ref{fig-cyberbullying} visualizes the number of state Cyberbullying laws enacted in a given year, alongside Cyberbullying news volume and mean article correlation.

\begin{table}[!htb]
\centering
\caption{Cyberbullying Laws by State. We omit the full list due to space constraints.}
\label{tab:cb}
\begin{tabular}{l r r }
\toprule
Year & State & Name\\
\midrule
2001 & CA &	SB719\\
2005 & AZ & HB2368\\
2006 & AK & HB482\\
2007 & AR &	Act115\\
2008 & CA & AB86\\
2009 & AL &	HB0216\\
2011 & AR &	Act905\\
2011 & AZ & HB2415\\
\multicolumn{3}{c}{Full list omitted}\\
2016 & KY & criminal statute 525.080\\
\bottomrule
\end{tabular}
\end{table}

\subsection{Drones}

\begin{figure}[!htb]
    \centering
    \resizebox{0.99\columnwidth}{!}{%
\begin{tikzpicture}
        \begin{axis}[
        title={},
        height=6cm,
        width=10cm,
        xlabel={Year},
        ylabel={Article Count},
        xmin=2003, xmax=2016,
        ymin=0, ymax=90,
        xtick={2003,2006,2009,2012,2015},
        ytick={0,10,20,30,40,50,60,70,80},
        legend pos=north west,
        legend style={font=\tiny},
        mark size=2.0pt,
        ymajorgrids=true,
        grid style=dashed,
        xticklabel style={/pgf/number format/.cd, set thousands separator={}},
        ]

        \addplot [very thick,blue,mark=o]
        coordinates {(2003,0) (2004,1) (2005,0) (2006,1) (2007,0) (2008,2) (2009,1) (2010,2) (2011,6) (2012,17) (2013,59) (2014,59) (2015,74) (2016,41)}
        node[left=.1cm,pos=0.7] {Articles};

    \end{axis}

\begin{axis}[
    height=6cm,
        width=10cm,
  axis y line*=right,
  axis x line=none,
  ymin=0, ymax=1,
xmin=2003, xmax=2016,
ytick={0,0.20,0.40,0.60,0.80,1.00},
  ylabel={Correlation and Sentiment}     
]

    \addplot [very thick,red,mark=x]
        coordinates {(2003,0) (2004,0) (2005,0) (2006,0.0322830578512397) (2007,0) (2008,0.0242122933884298) (2009,3.22830578512397e-04) (2010,0.00343677405445009) (2011,0.00942874162706814) (2012,0.230384688573092) (2013,0.231828700724021) (2014,0.263146275804372) (2015,0.292283420413794) (2016,0.116639219043336)}
        node[below right=.05cm,pos=0.65] {Correlation};

    \addplot [very thick,black,mark=square]
        coordinates {(2003,0) (2004,0.48) (2005,0) (2006,0.49) (2007,0) (2008,0.46) (2009,0) (2010,0.41) (2011,0.47) (2012,0.36) (2013,0.27) (2014,0.23) (2015,0.21) (2016,0.39)}
        node[above=.1cm,pos=0.5] {Sentiment};
	
    \end{axis}
\end{tikzpicture}%
}
    \caption{News volume and correlation as predictors of legislation in Drones.}
    \label{fig-drones}
\end{figure}
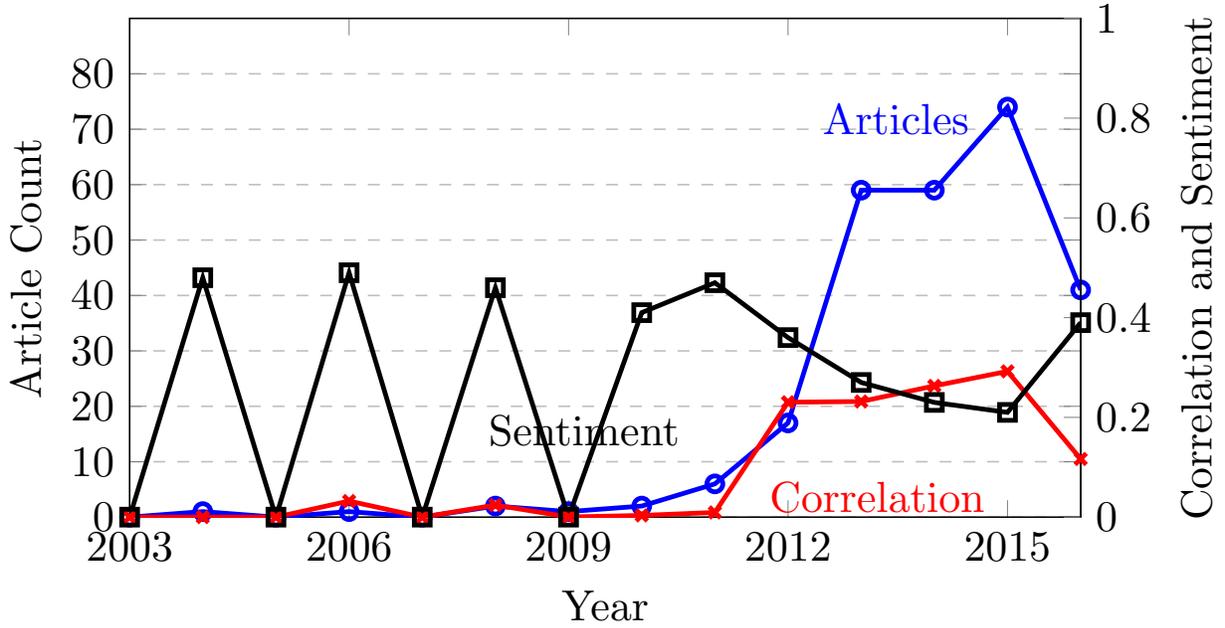

Drone legislation in America was first promulgated in 2015 \citep{dl}, and Senate debate on the subject has been active since. We tested our approach on this dataset by using the data shown in Figure~\ref{fig-drones}. As can be seen from the figure, Volume and MNC reach a peak for the year 2015, while sentiment is at a low for the period surveyed (we exclude years in which the Volume is 0). We tested our classifier on 8 labels from this set, excluding years prior to 2009 due to the absence of relevant publishing during those years (Figure~\ref{fig-drones}). We achieve an F-measure of 0.875 on this topic.

\subsection{Child Privacy}
\begin{figure}[!htb]
    \centering
    \resizebox{0.99\columnwidth}{!}{%
\begin{tikzpicture}
        \begin{axis}[
        title={},
        height=6cm,
        width=10cm,
        xlabel={Year},
        ylabel={Article and Law Count},
        xmin=1991, xmax=2016,
        ymin=0, ymax=200,
        xtick={1990,1995,2000,2005,2010,2015},
        ytick={0,30,60,90,120,150,180},
        legend pos=north west,
        legend style={font=\tiny},
        mark size=2.0pt,
        ymajorgrids=true,
        grid style=dashed,
        xticklabel style={/pgf/number format/.cd, set thousands separator={}},
        ]

        \addplot [very thick,blue,mark=o]
        coordinates {(1991,30) (1992,24) (1993,39) (1994,30) (1995,23) (1996,24) (1997,42) (1998,54) (1999,67) (2000,142) (2001,165) (2002,121) (2003,162) (2004,166) (2005,134) (2006,177) (2007,85) (2008,67) (2009,34) (2010,41) (2011,36) (2012,136) (2013,103) (2014,87) (2015,66) (2016,56)}
        node[left=.6cm, pos=0.3] {Articles};
        
       \addplot [very thick,brown,mark=square]
        coordinates {(1991,0) (1992,0) (1993,0) (1994,0) (1995,0) (1996,0) (1997,0) (1998,0) (1999,0) (2000,100) (2001,100) (2002,100) (2003,100) (2004,100) (2005,100) (2006,100) (2007,0) (2008,0) (2009,0) (2010,0) (2011,0) (2012,0) (2013,100) (2014,100) (2015,0) (2016,0)}
        node[below=.1cm, pos=0.264] {COPPA};

    \end{axis}

\begin{axis}[
    height=6cm,
        width=10cm,
  axis y line*=right,
  axis x line=none,
xmin=1991, xmax=2016,
  ymin=0, ymax=1,
ytick={0,0.20,0.40,0.60,0.80,1.00},
  ylabel={Correlation and Sentiment}     
]

    \addplot [very thick,red,mark=x]
        coordinates {(1991,0.494891730377283) (1992,0.306500824918334) (1993,0.450524639496613) (1994,0.244670987124601) (1995,0.500434273364260) (1996,0.167088600014157) (1997,0.234390727887453) (1998,0.313505617206397) (1999,0) (2000,0.950829151665395) (2001,0.836356036667068) (2002,0.524935759360227) (2003,0.9234556) (2004,0.9502805248098893) (2005,0.672068049974419) (2006,0.778970431475917) (2007,0.593680956199340) (2008,0.538252502878553) (2009,0.243932145909111) (2010,0.403037597968340) (2011,0.365651526803978) (2012,0.798736564775931) (2013,0.775975894472513) (2014,0.730283954635817) (2015,0.340809422941126) (2016,0.466103653044484)}
        node[above=0.1cm, pos=0.85] {Correlation};

           \addplot [very thick,black,mark=square]
        coordinates {(1991,.61) (1992,.63) (1993,.62) (1994,.66) (1995,.57) (1996,.61) (1997,.63) (1998,.54) (1999,.52) (2000,.47) (2001,.64) (2002,.49) (2003,.45) (2004,.47) (2005,.46) (2006,.67) (2007,.63) (2008,.67) (2009,.49) (2010,.41) (2011,.36) (2012,.57) (2013,.53) (2014,.52) (2015,.54) (2016,.56)}
        node[above=.2cm,pos=0.1] {Sentiment};

    \end{axis}

\end{tikzpicture}%
}
    \caption{News volume and correlation as predictors of legislation in child privacy.}
    \label{fig-child}
\end{figure}
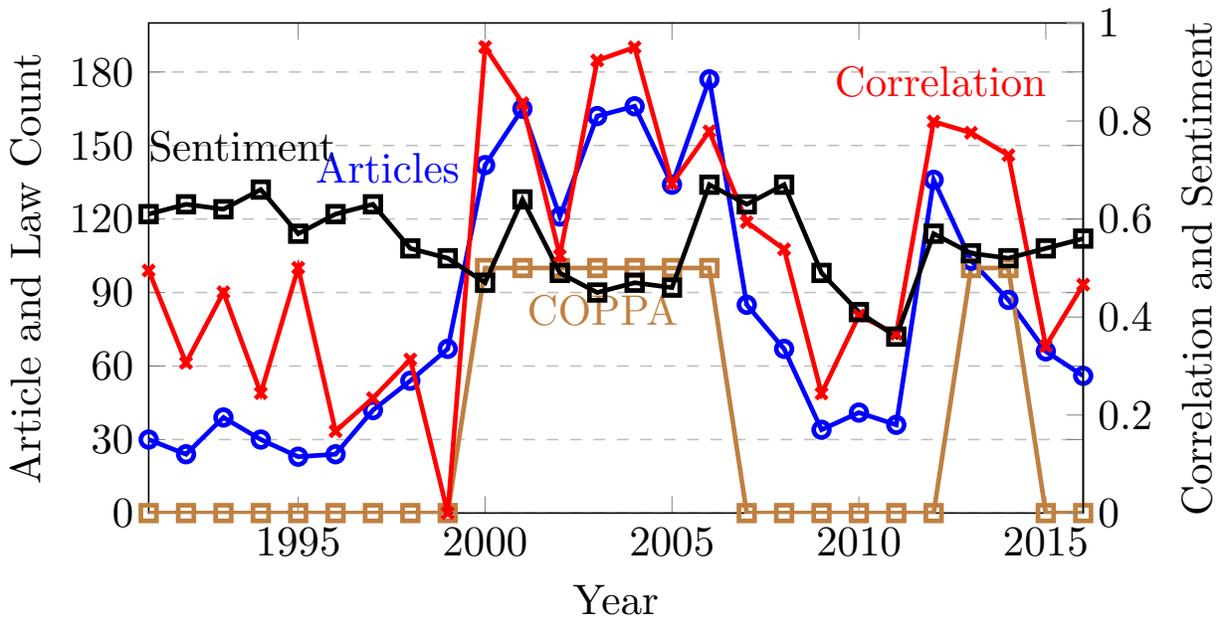

The primary laws governing children's privacy protection in the United States are COPPA \citep{coppa} and FERPA \citep{ferpa}. COPPA was originally introduced in April 1998, and went through a series of amendments from 1999 through 2005, and again from 2012--2013. FERPA was enacted in 1974. Due to the unavailability of children's privacy news articles before 1974 (a keyword search in the NYT developers API returns 0 articles), we restrict our analysis to COPPA. We collected children's privacy news articles from 1990 to 2016 from the NYT API (a total of 2,011 articles), and visualize news volume together with correlation in Figure~\ref{fig-child}. The figure displays a clear correlation between news volume and likelihood of legislation. The LOO approach produces an F-measure of 1 on this dataset.




\subsection{HTML5}

\begin{figure}[!htb]
    \centering
    \resizebox{0.99\columnwidth}{!}{%
\begin{tikzpicture}
        \begin{axis}[
        title={},
        height=6cm,
        width=10cm,
        xlabel={Year},
        ylabel={Article Count},
        xmin=2008, xmax=2014,
        ymin=0, ymax=50,
        xtick={2008,2009,2010,2011,2012,2013,2014},
        ytick={0,10,20,30,40,50},
        legend pos=north west,
        legend style={font=\tiny},
        mark size=2.0pt,
        ymajorgrids=true,
        grid style=dashed,
        xticklabel style={/pgf/number format/.cd, set thousands separator={}},
        ]

        \addplot [very thick,blue,mark=o]
        coordinates {(2008,0) (2009,2) (2010,41) (2011,28) (2012,24) (2013,9) (2014,2)}
        node[above left,pos=0.6] {Articles};

    \end{axis}

\begin{axis}[
    height=6cm,
        width=10cm,
  axis y line*=right,
  axis x line=none,
 xmin=2008, xmax=2014,
  ymin=0, ymax=1,
ytick={0,0.20,0.40,0.60,0.80,1.00},
  ylabel={Correlation and Sentiment}     
]

   \addplot [very thick,red,mark=x]
        coordinates {(2008,0) (2009,.10127664) (2010,.26773982) (2011,.16334521) (2012,.13433216) (2013,.12215464) (2014,.10436654) }
        node[below right,pos=.65] {Correlation};

    \addplot [very thick,black,mark=square]
        coordinates {(2008,0) (2009,.4632) (2010,0.1366) (2011,0.1574) (2012,.2271) (2013,.3077) (2014,.2939)}
        node[above=.1cm,pos=0.5] {Sentiment};

    \end{axis}
\end{tikzpicture}%
}
    \caption{News patterns around the 2011 Facebook ID leak, and subsequent legislation.}
    \label{fig-html5}
\end{figure}
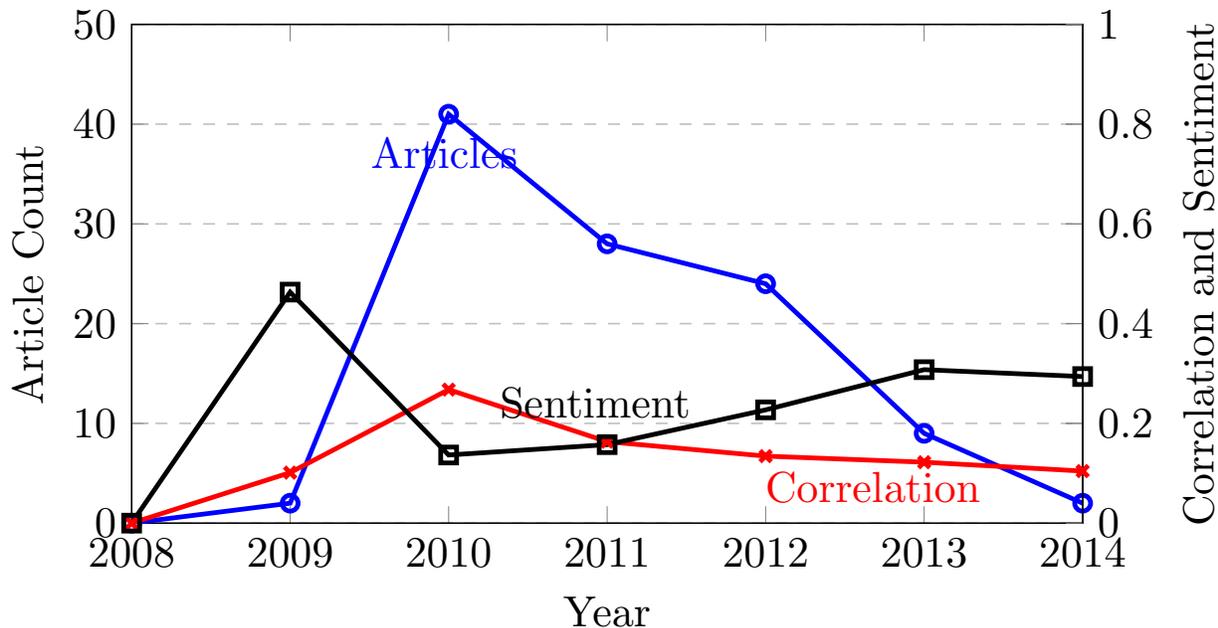

Figure~\ref{fig-html5} depicts the news patterns for the HTML5 Facebook ID leak case, from a quiescent period (2008) to the period capturing the leak, and subsequent legislative reaction (2011--2014). Our model correctly predicts five out of the seven labels. 

\section{Limitations}
\mps{Broken para}
Our analysis leaves out \emph{social media}, ex: twitter, and focuses on just two (albeit influential) news sources. While sufficiently predictive for our dataset, our model would benefit from additional data sources. The dependence on earlier surveys to extract positives limited the cardinality of our dataset, however, our approach remains generic. 

\section{Conclusion}
We highlight an influential facet of news, framing, which has hitherto been ignored by the computer science community. We demonstrate that existing approaches (TDT, DLDA) fail to detect framing changes. We contribute a simple entropic algorithm that together with learned semantic similarity detects framing changes with high precision. Further, we posit a counter intuitive relationship between the volume of topic based publishing and the similarity of the published articles. This enables us to estimate the nature of a current topic news cycle, how long it is likely to endure, and the nature of likely public reaction. We demonstrate the practical utility of our approach with a case study of legislation in topics of current interest, and achieve an average F-measure of 0.96 on our dataset. Our work demonstrates for the first time that topic news patterns have predictive utility.



\cwh{References have improved, but there are some mistakes.  For example, \citep{hu-liu} is a conference paper not a journal.}

\bibliographystyle{aaai}

\bibliographystyle{aaai}


\end{document}